\documentclass[conference]{IEEEtran}
\IEEEoverridecommandlockouts
\usepackage{cite}
\usepackage{amsmath,amssymb,amsfonts}
\usepackage{hyperref}
\usepackage{float}
\usepackage{algorithm,algcompatible}
\DeclareMathOperator*{\argmin}{argmin}
\algnewcommand\INPUT{\item[\textbf{Input:}]}%
\algnewcommand\OUTPUT{\item[\textbf{Output:}]}%

\usepackage{graphicx}
\usepackage{textcomp}
\usepackage{xcolor}
\def\BibTeX{{\rm B\kern-.05em{\sc i\kern-.025em b}\kern-.08em
    T\kern-.1667em\lower.7ex\hbox{E}\kern-.125emX}}

\usepackage{tikz}
\usepackage{xcolor}
\usepackage{pgfplots}
\usetikzlibrary{spy}
\pgfplotsset{compat=1.18} 
\begin{document}

\thispagestyle{empty}
\pagestyle{empty}
\title{Self-Supervised Hyperspectral Inpainting with the Optimisation inspired Deep Neural Network Prior\\

}

\author{\IEEEauthorblockN{Shuo Li}
\IEEEauthorblockA{\textit{School of Engineering} \\
\textit{University of Edinburgh}\\
s1809498@ed.ac.uk}
\and
\IEEEauthorblockN{Mehrdad Yaghoobi}
\textit{University of Edinburgh}\\
m.yaghoobi-vaighan@ed.ac.uk}

\maketitle
\begin{abstract}
Hyperspectral Image (HSI)s cover hundreds or thousands of narrow spectral bands, conveying a wealth of spatial and spectral information. However, due to the instrumental errors and the atmospheric changes, the HSI obtained in practice are often contaminated by noise and dead pixels(lines), resulting in missing information that may severely compromise the subsequent applications. We introduce here a novel HSI missing pixel prediction algorithm, called Low Rank and Sparsity Constraint Plug-and-Play (LRS-PnP). It is shown that LRS-PnP is able to predict missing pixels and bands even when all spectral bands of the image are missing. The proposed LRS-PnP algorithm is further extended to a self-supervised model by combining the LRS-PnP with the Deep Image Prior (DIP), called LRS-PnP-DIP. In a series of experiments with real data, It is shown that the LRS-PnP-DIP either achieves state-of-the-art inpainting performance compared to other learning-based methods, or outperforms them. 
\end{abstract}


\section{Introduction}
In hyperspectral imagery, sensor failures and malfunctions of the HSI acquisition system may result in missing pixels/lines or some spectral bands, significantly hindering the subsequent processing of observed HSI \cite{roi2012anisotropic}. Hyperspectral inpainting is the task of filling in the missing areas with plausible contents. However, the inpainting of HSIs is a more challenging task than RGB images as each pixel to be filled in is a complex vector with rich spatio-spectral information. Traditional methods such as \cite{zhuang2018fast}, \cite{LRTV}, \cite{sparse_representation} and \cite{low_rank} either fail when the whole spectral bands of pixels are missing, or their performance is severely compromised if there are a large number of missing pixels. In this work, we treat hyperspectral inpainting as a special case of the reconstruction problem, whose objective is to recover the ground truth from the degraded/masked incomplete images. The low rankness and sparsity of the underlying clean HSI are used here as the priors during reconstruction. The sparse representation (SR) and low rankness (LR) priors have been successfully applied in a wide range of hyperspectral imaging applications such as classification \cite{SR-application_1}, denoising \cite{SR-application_2}, and un-mixing\cite{SR-application_3}.
Recently, researchers have found that the missing spectrum of the HSIs can be predicted through learning on a large dataset \cite{wong2020hsi} or learning from the image itself \cite{sidorov2019deep}. The latter is an extension of the Deep Image Prior (DIP)\cite{ulyanov2018deep} applied to HSIs, achieving the state-of-the-art performance. In \cite{ulyanov2018deep}, authors reveal that the structure of a generative network is sufficient to capture plenty of low-level image statistics prior to any learning. The well-designed CNN networks have been tested on a wide range of tasks such as image denoising, super-resolution and image inpainting, showing promising results which are even competitive to the state-of-the-art deep models trained on large datasets. The ``free of external training data" property of the DIP, makes it well-suited for HSI inpainting. The very recent works in \cite{2021_DIP_In_Loop,wu2022adaptive} raised the point that some trained or untrained neural networks can be directly plugged into the iterative solver to achieve better reconstruction accuracy. Keeping this Plug-and-Play (PnP) idea in mind, one may design a better HSI inpainting algorithm by taking advantage of both the traditional and deep learning techniques. Considering the high computational cost of end-to-end training on extensive HSI data, it is here proposed to use DIP, a self-supervised framework that is free of external training data.

\subsection{Contribution}
This paper aims to develop an effective HSI inpainting algorithm that enjoys the specific learning capability of deep networks, called inductive bias, but does not need any external training data, i,e. self-supervised learning. Speciﬁc contributions are the following:
\begin{itemize}
\item A novel self-supervised HSI inpainting method, called Low Rank and Sparsity Constraint Plug and play (LRS-PnP), which is presented to solve the most challenging scenarios where the whole spectral bands are missing. 
\item The use of deep neural networks in replacing the rank constrained-optimisation problem, showing the potential of DIP in learning intrinsic low-rank characteristics of the HSIs.
\item A deep hyperspectral prior-based model which better exploits the intrinsic characteristics of HSI data, for achieving state-of-the-art performance.
\item Extensive experiments on real data to verify the superiority of proposed LRS-PnP and LRS-PnP-DIP algorithms over existing inpainting solutions.
\end{itemize}
The rest of paper is organized as follows: section \textrm{II} introduces the proposed framework, section \textrm{III} provides the implementation details and experimental setups, in section \textrm{IV}, the results are discussed. Finally, section \textrm{V} concludes the paper.
\section{Proposed Method}
The HSI inpainting task can be interpreted as the reconstruction of the clean image $X$ from its noisy and incomplete measurement $Y$, in the presence of additive noise $N$ and masking operator $M$:
\begin{equation}
 Y = M\{X\} +N
\end{equation}
The clean image $X \in\mathbb{R}^{q}$($q =n_{r} \times n_{c} \times n_{b}$), where $n_{r},n_{c}$ are the spatial dimension of image, and $n_b$ stands for the total numbers of spectral bands. The operator $M: \mathbb{R}^{q} \rightarrow \mathbb{R}^{q}$ is the binary mask, with 0 representing the missing pixel and 1 representing the observed and valid pixel. Therefore $M$ can be presented with a diagonal matrix that has only one per row, which we note it by $\rm M$. $N$ is the additive Gaussian noise of appropriate size. Specifically, $M$ is often given. The formulation (1) is a linear system which can be written as:\\ 
\begin{equation}
 \boldsymbol{y} = \rm M \boldsymbol{x} +\boldsymbol{n}
\end{equation}
Where $\boldsymbol{x},\boldsymbol{y},\boldsymbol{n}$ are the vectorized forms of $X,Y,N$, respectively, and $\rm M$ is a diagonal matrix.
We first introduce an operator $P_i(\boldsymbol{x})$ that extracts each i-th patch from the image $\boldsymbol{x}$. $P_i(\boldsymbol{x})$ may cover only the valid pixels or may include the missing pixels depending on the size of $P_i(\cdot)$. We apply the sparse representation on each image patch $P_i(\boldsymbol{x})$. The inpainted image $\boldsymbol{x^*}$ can be obtained by solving the following optimization problem:
\begin{equation}
\begin{aligned}
 (\boldsymbol{x^*},\boldsymbol{\alpha^*}) = &\argmin_{\boldsymbol{x}, \boldsymbol{\alpha}}   \gamma\Vert \boldsymbol{y} -\rm M\boldsymbol{x} \Vert_{2}^2 + w_{lr}\Vert \boldsymbol{x} \Vert_* + w_{s}\Vert \boldsymbol{\alpha} \Vert_1\\
&\textrm{s.t.} \quad \boldsymbol{x} = \Phi \boldsymbol{\alpha} 
\end{aligned}
\end{equation}
The first term is the data fidelity term, which we weigh with the parameter $\gamma$. Due to the ill-posed nature of estimating $\boldsymbol{x}$ from $\boldsymbol{y}$ only using data fidelity term, the solution is often not unique. For this reason, we introduce another two "priors" to regularize the inpainting problem, namely low rank and sparsity constraints. The second term penalizes the solution $\boldsymbol{x}$ to be of low rank, which is often used as the surrogate for the rank minimization problem. The third term constrains the missing pixels to be generated from the subspace approximated by the valid pixels. Similarly, we weigh these two terms with parameters $w_{lr}$ and $w_{s}$. The sparse representation problem is solved with a known dictionary $\Phi$ that is learned only from the noisy and incomplete pixels, or it is a sparsifying transform.
By adopting the augmented Lagrangian and introducing the auxiliary variable $\boldsymbol{u}$\cite{ADMM}, problem (2) can be rewritten as:
\begin{equation}
\begin{aligned}
(\boldsymbol{x^*},\boldsymbol{\alpha^*}) = &\argmin_{\boldsymbol{x},\boldsymbol{\alpha}} \gamma \Vert \boldsymbol{y} -\rm M\boldsymbol{x} \Vert_{2}^2 + w_{lr}\Vert \boldsymbol{u} \Vert_*
 + w_{s}\sum_i \Vert \boldsymbol{\alpha}_i \Vert_1. \\
 &+\frac{\boldsymbol{\mu}_1}{2}\Vert \sum_i(P_i(\boldsymbol{x})-\Phi \boldsymbol{\alpha}_i) +\frac{\boldsymbol{\lambda}_1}{\boldsymbol{\mu}_1} \Vert_{2}^2\\
&\textrm{s.t.} \quad \boldsymbol{x} = \boldsymbol{u}
\end{aligned}
\end{equation}
Where $\boldsymbol{\lambda}_1$ and $\boldsymbol{\mu}_1$ are the Lagrangian multiplier and penalty term, respectively. With the help of the alternating direction method of multipliers (ADMM), problem (4) can be solved by the sequential updates of 
three variables: $\boldsymbol{\alpha}$, $\boldsymbol{u}$ and $\boldsymbol{x}$. \\
1) \textit{Fixing $\boldsymbol{u}$ and $\boldsymbol{x}$, and updating $\boldsymbol{\alpha}$}:
\begin{equation}
\begin{aligned}
 {\boldsymbol{\alpha}}^{k+1} = \argmin_{\boldsymbol{\alpha}} \frac{\boldsymbol{\mu}_1^k}{2} \sum_i\Vert(P_i(\boldsymbol{x}^k) + \frac{\boldsymbol{\lambda}_1^k}{\boldsymbol{\mu}_1^k})-\Phi \boldsymbol{\alpha}_i \Vert_{2}^2 \\+  w_{s}\sum_i \Vert \boldsymbol{\alpha}_i \Vert_1 \\
\end{aligned}
\end{equation}
which is a patched-based sparse coding problem which can be solved  using iterative solvers. In our algorithm, we adopt the PnP-ISTA \cite{PnP-ISTA}, which has shown promising results over conventional ISTA\cite{ISTA}. Denote the first term in equation (5) as $f = \frac{\boldsymbol{\mu}_1^k}{2} \sum_i\Vert(P_i(\boldsymbol{x}^k) + \frac{\boldsymbol{\lambda}_1^k}{\boldsymbol{\mu}_1^k})-\Phi \boldsymbol{\alpha}_i \Vert_{2}^2$. The whole process can then be replaced by an off-the-shelf denoiser $\mathcal{D}$ acting on the gradient of $f$, as it is proposed in \cite{PnP-ISTA}. Every single iterate takes the form: \\
\begin{equation}
\begin{aligned}
 {\boldsymbol{\alpha}^{k+1}} = \mathcal{D}(I-\nabla f)(\boldsymbol{\alpha}^{k})
\end{aligned}
\end{equation}
2)\textit{Fixing $\boldsymbol{\alpha}$ and $\boldsymbol{x}$, and updating $\boldsymbol{u}$}:
\begin{equation}
\begin{aligned}
  {\boldsymbol{u}^{k+1}} = \argmin_{\boldsymbol{u}} w_{lr}\Vert \boldsymbol{u} \Vert_* + \frac{\boldsymbol{\mu_2}^{k}}{2}\Vert (\boldsymbol{x}^k+\frac{\boldsymbol{\lambda_2}^k}{\boldsymbol{\mu}_2^k}) - \boldsymbol{u} \Vert_{2}^2
\end{aligned}
\end{equation}
which can be solved by the Singular Value Thresholding (SVT) algorithm. The element-wise soft shrinkage is applied to the singular value of $(\boldsymbol{x}^k + \frac{\boldsymbol{\lambda}_2^k}{\boldsymbol{\mu}_2^k})$, as follows,
\begin{equation}
\begin{aligned}
  {\boldsymbol{u}^{k+1}} = SVT(\boldsymbol{x}^k+\frac{\boldsymbol{\lambda}_2^k}{\boldsymbol{\mu}_2^k})
\end{aligned}
\end{equation}
In the proposed LRS-PnP-DIP algorithm, update step of (8) is replaced by a untrained randomised weight neural network $f_\theta(\boldsymbol{z})$, where $\theta$ represents the network weights to be updated, and the input $\boldsymbol{z}$ is set to be $\boldsymbol{x}^k + \frac{\boldsymbol{\lambda}_2^k}{\boldsymbol{\mu}_2^k}$. i,e, the latent image from the previous iterations: \\
\begin{equation}
\begin{aligned}
 {\boldsymbol{u}^{k+1}} = f_\theta(\boldsymbol{x}^k + \frac{\boldsymbol{\lambda}_2^k}{\boldsymbol{\mu}_2^k})
\end{aligned}
\end{equation} \\
3) \textit{Fixing $\alpha$ and $\boldsymbol{u}$, and updating $\boldsymbol{x}$}: \\
\begin{equation}
\begin{aligned}
  {\boldsymbol{x}^{k+1}} = \argmin_{\boldsymbol{x}}  \gamma\Vert \boldsymbol{y} -\rm M\boldsymbol{x} \Vert_{2}^2 + \sum_i\Vert(P_i(\boldsymbol{x}) + \frac{\boldsymbol{\lambda}_1^k}{\boldsymbol{\mu}_1^k})-\Phi\boldsymbol{\alpha}_i^{k+1} \Vert_{2}^2 \\
  + \frac{\boldsymbol{\mu}_2^k}{2}\Vert (\boldsymbol{x}+\frac{\boldsymbol{\lambda}_2^k}{\boldsymbol{\mu}_2^k}) - \boldsymbol{u}^{k+1} \Vert_{2}^2 \\
\end{aligned}
\end{equation}
Closed-form solution for $\boldsymbol{x}$ exists as follows: \\
\begin{equation}
\begin{aligned}
  &{\boldsymbol{x}^{k+1}} = (\gamma \rm M^T\rm M + \boldsymbol{\mu}_1^k \sum_i P_i^TP_i        +\boldsymbol{\mu}_2^k\rm I)^{-1} \\ 
   &( \gamma \rm M^T\boldsymbol{y}+ \boldsymbol{\mu}_1^k \sum_i P_i\Phi \boldsymbol{\alpha}_i^{k+1} + \boldsymbol{\mu}_2^k \boldsymbol{u}^{k+1} - \sum_i P_i \boldsymbol{\lambda}_1^k -  \boldsymbol{\lambda}_2^k \rm I) \\
\end{aligned}
\end{equation}
4) \textit{Lagrangian and penalty terms updating}: \\
\begin{equation}
\begin{aligned}
  {\boldsymbol{\lambda}_1^{k+1}} = \boldsymbol{\lambda}_1^{k} + \boldsymbol{\mu}_1^{k}(\boldsymbol{x}^{k+1}-\Phi \boldsymbol{\alpha}^{k+1})\\
   {\boldsymbol{\lambda}_2^{k+1}} = \boldsymbol{\lambda}_2^{k} + \boldsymbol{\mu}_2^{k}(\boldsymbol{x}^{k+1}-\boldsymbol{u}^{k+1})
\end{aligned}
\end{equation}
\begin{equation}
\begin{aligned}
 \boldsymbol{\mu}_1^{k+1} = \boldsymbol{\rho}_1 \boldsymbol{\mu}_1^{k}\\
 \boldsymbol{\mu}_2^{k+1} = \boldsymbol{\rho}_2 \boldsymbol{\mu}_2^{k}
\end{aligned}
\end{equation} \\
The proposed Low-Rank and Sparsity Plug-and-Play (LRS-PnP) inpainting model is presented in Algorithm 1. 
\begin{algorithm}
    \caption{(LRS-PnP) Algorithm}
  \begin{algorithmic}[1]
    \REQUIRE masking matrix: $\rm M$, noisy and incomplete HSI: $\boldsymbol{y}$, learned dictionary: $\Phi$. denoiser: $\mathcal{D}$, max iteration: $It_{max}$.
    
    \OUTPUT inpainted HSI image $X$.
    \STATE \textbf{Initialization}: $\boldsymbol{\lambda}_1,\boldsymbol{\lambda}_2,\boldsymbol{\mu}_1,\boldsymbol{\mu}_2, \boldsymbol{\rho}_1,\boldsymbol{\rho}_2$.
      \WHILE{Not Converged}
        \STATE  for $i=1:It_{max}$ do:

               $\boldsymbol{\alpha}^{k+1} = \mathcal{D}(I-\nabla f)(\boldsymbol{\alpha}^{k})$
      
         \STATE ${\boldsymbol{u}^{k+1}} = SVT(\boldsymbol{x}^k+\frac{\boldsymbol{\lambda}_2^k}{\boldsymbol{\mu}_2^k})$
         \STATE update $\boldsymbol{x}$ by ((11)).
         \STATE update Lagrangian parameters and penalty terms.
    \ENDWHILE
  \end{algorithmic}
\end{algorithm} \\
By replacing the SVT with DIP $f_{\theta}$, we end up with an extension of the LRS algorithm, denote as LRS-PnP-DIP, is presented in Algorithm 2:
\begin{algorithm}[H]
    \caption{(LRS-PnP-DIP) Algorithm}
  \begin{algorithmic}[1]
    \REQUIRE masking matrix: $\rm M$, noisy and incomplete HSI: $\boldsymbol{y}$, learned dictionary: $\Phi$. denoiser: $\mathcal{D}$, max iteration: $It_{max}$. DIP: $f_{\theta}$
    \OUTPUT inpainted HSI image $\boldsymbol{x}$.
    \STATE \textbf{Initialization} DIP parameters, $\boldsymbol{\lambda}_1,\boldsymbol{\lambda}_2,\boldsymbol{\mu}_1,\boldsymbol{\mu}_2, \boldsymbol{\rho}_1,\boldsymbol{\rho}_2$.
    \WHILE{Not Converged}
      \STATE  for $i=1:It_{max}$ do:
      
               $\boldsymbol{\alpha}^{k+1} = \mathcal{D}(I-\nabla f)(\boldsymbol{\alpha}^{k})$
      \STATE update $\theta$ in DIP, with the target $\boldsymbol{y}$ and input $\boldsymbol{x}^k + \frac{\boldsymbol{\lambda}_2^k}{\boldsymbol{\mu}_2^k}$.
      \STATE update $\boldsymbol{x}$ by (11).
      \STATE update Lagrangian parameters and penalty terms.
    \ENDWHILE
  \end{algorithmic}
\end{algorithm}
\section{Implementation Details}
We test the proposed inpainting model on the Chikusei airborne hyperspectral dataset, which was taken by Headwall Hyperspec-VNIR-C imaging sensor \cite{Chikusei}. The hyperspectral test image has 128 spectral bands, with each 36x36 pixels. The size of the dictionary $\Phi$ is 1296x2000 which was learned only based on the noisy and incomplete HS image. All images are corrupted with Gaussian noise with the fixed noise strength $\sigma=0.12$. The mask $\rm M$ is applied to all the spectral bands. We use BM3D and Non-local-Mean (NLM) as the plug-and-play denoiser. The implementation of DIP follows the same structures as in Deep Hyperspectral Prior paper\cite{sidorov2019deep}. To prevent over-ﬁtting in DIP, we use the early stopping criterion proposed in \cite{wang2021early}, which automatically detects the near-peak PSNR point using windowed moving variance (WMV). We set $w_{s}/w_{lr}$ to be 1 and $\gamma$ to be 0.5. Note that the choice of $\gamma$ is highly related to the noise level of $Y$. If the noise level is low, the recovered image $X$
should be close to noisy observation $Y$, then parameter $\gamma$ should be
large, and vice versa. For ADMM parameters, $\lambda_1$ and $\lambda_2$ are set to be 0, $\mu_1$, $\mu_2$ are set to be 1, and $\rho_1$, $\rho_2$ are set to be 1, meaning a fixed update step. We use Adam optimizer, and the learning rate is set to be 0.1. We adopt two widely used indicators: Mean Signal-to-Noise Ratio (MPSNR) and Mean Structural Similarity (MSSIM), to evaluate the performance in all experiments.

\section{Experimental Results}
The performance of proposed algorithms are compared with the existing traditional methods LRTV\cite{LRTV}, and FastHyIn\cite{zhuang2018fast}, and learning-based method DIP\cite{ulyanov2018deep}, DeepRED\cite{2019_Deep_Red} and PnP-DIP\cite{pnp_dip}. The numerical results are reported in Table $I$ and $II$. For the DIP, DeepRED, PnP-DIP and the proposed LRS-PnP-DIP, the same U-net in \cite{sidorov2019deep} was used as the backbone. For the fairness of comparison, all associated parameters with U-net are kept fixed, including the noise standard deviations and regularization strength. The MPSNR and MSSIM are obtained through running algorithms for 20 times on the same test image. For comparison with FastHyIn and LRTV, 25\% of the pixels were randomly masked using a uniform distribution for the missing bands. The inpainting performance is shown in Figure 1.
\begin{table}[H]
\centering
\begin{tabular}{c rrrrr}
\hline\hline
Methods & Input & LRTV\cite{LRTV} & FastHyIn\cite{zhuang2018fast} & LRS-PnP\\
\hline 
MPSNR$\uparrow$ & 32.745 & 37.655 & 40.876 & $\boldsymbol{41.263}$\\ 
\hline 
MSSIM $\uparrow$ & 0.292 & 0.720 & 0.902 & $\boldsymbol{0.935}$\\ 
\hline
\end{tabular}
\label{tab:hresult}
\caption{Performance Comparing with Traditional Methods} 
\end{table} 
It shows that LRS-PnP algorithm can generate a more consistent and realistic spectrum in the missing region compared to the learning-based methods. LRS-PnP captures local structures, such as a sudden change in the materials, e,g. see the first image. In contrast, methods such as DeepRED and PnP-DIP tend to generate much smoother contents in the non-missing areas. However, there are severe distortions and artefacts of the missing regions. We believe this is due to the nature of DIP which mainly focuses on learning global features and characteristics. For this reason, it is often used with other regularizers such as Total Variation (TV) to preserve more details\cite{2019_DIP_TV}.
It is worth mentioning that the inpainting result of the double deep prior algorithm LRS-PnP-DIP is visually and qualitatively better than those of other inpainters.

\begin{table*}
\centering
\begin{tabular}{c rrrrrr}
\hline\hline
Methods & Input & DIP\cite{ulyanov2018deep} & DeepRED\cite{2019_Deep_Red}  &PnP-DIP\cite{pnp_dip} & LRS-PnP(Ours) & LRS-PnP-DIP(Ours)\\
\hline 
MPSNR $\uparrow$ & 31.569 & 41.247($\pm$ 0.62)  & 41.35($\pm$ 0.32) & 41.52($\pm$ 0.35)  & $\boldsymbol{40.802}$& $\boldsymbol{42.385(\pm0.28)}$\\ 
\hline 
MSSIM $\uparrow$ & 0.268 & 0.937($\pm$ 0.004) & 0.942($\pm$ 0.001)   & 0.947($\pm$ 0.002) & $\boldsymbol{0.918}$& $\boldsymbol{0.954(\pm0.002)}$\\ 
\hline
\end{tabular}

\caption{Performance Comparing with Learning-Based Methods. The average and variance over 20 sample generation is shown here} 
\end{table*} 
\begin{figure*}    
  \scalebox{0.95}{
  \hspace{0.4cm}
  \includegraphics[width=1.0\textwidth,height=0.55\textwidth,]{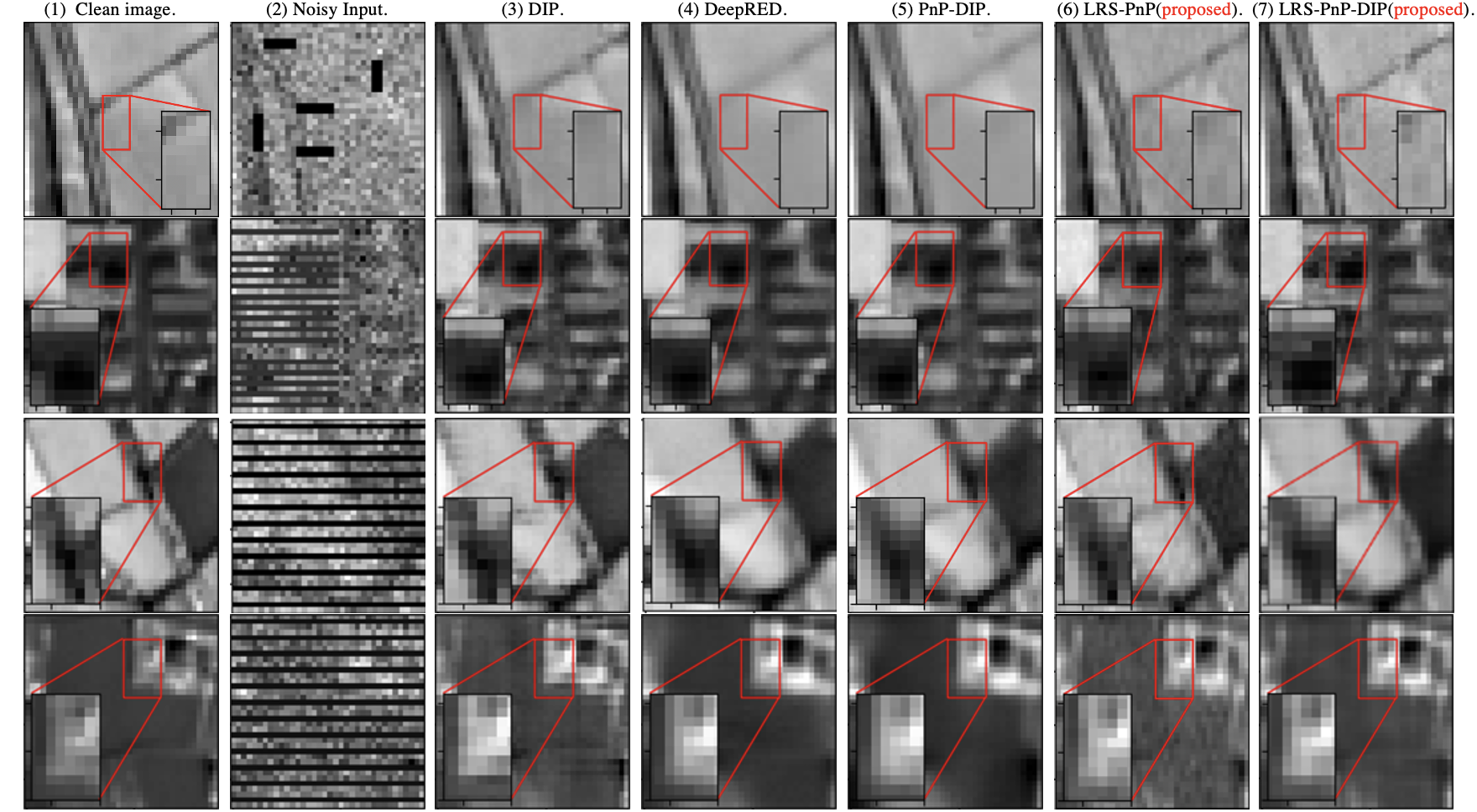}
 }
   \vspace{-0.3cm}
  \caption{Comparison of our solution and other learning-based inpainting algorithms. From left to right: (1) clean image, (2) input image, (3) Deep Image Prior, (4) DeepRED, (5) PnP-DIP, (6) proposed LRS-PnP, and (7) proposed LRS-PnP-DIP. All images are visualized at band 80.}
  \label{img}  
\end{figure*} 

\section{Conclusion}
The novel hyperspectral inpainting algorithms called LRS-PnP and its extension self-supervised LRS-PnP-DIP which can effectively handle missing pixels from noisy and incomplete HS images in the most challenging scenario where the whole spectrum bands are missing. The new methods exploit spectral and spatial redundancy of HSIs and require no training data except, the test image. A comparison of LRS-PnP and LRS-PnP-DIP with the state-of-the-art algorithms is conducted on a real HS image, leading to the conclusion that LRS-PnP yields similar while LRS-PnP-DIP yields better performance against other learning-based methods, while not using a pre-training step using a large set of HS images. As the future work, one direction is to accelerate the proposed algorithms to achieve a cost-efficient and real-time HSI inpainting. Another direction of interest is to explore and optimize the training of DIP, especially when it is used in the loop, as mentioned in a lines of DIP-related works\cite{2021_DIP_In_Loop,pnp_dip,DIP-TV,lai2022deep,wu2022adaptive}. The other direction is to theoretically show the convergence of the algorithm, which we left for a future work.

\bibliographystyle{unsrt}

\bibliography{Bibliography} 
\vspace{12pt}
\color{red}
\end{document}